\documentclass[11pt]{article}

\usepackage{deauthor,times,graphicx}
\usepackage{xspace}
\usepackage{xcolor}

\usepackage{url}

\usepackage{subcaption}
\usepackage{float}
\usepackage{caption}

\newcommand{\code}[1]{\texttt{#1}}
\newcommand{\topic}[1]{\vspace{0.25em}\noindent\textbf{#1.}}

\definecolor{ao}{rgb}{0.0, 0.5, 0.0}

\newcommand{\stitle}[1]{\noindent{\bf #1}}

\newcommand{\blah}[1]{\textcolor{red}{blah}}

\newcommand{\thinktime}[0]{think time\xspace}
\newcommand{\Thinktime}[0]{Think time\xspace}

\begin{document}

\title{Enhancing the Interactivity of Dataframe Queries \\ by Leveraging Think Time}

\author{Doris Xin, Devin Petersohn, Dixin Tang, Yifan Wu, Joseph E. Gonzalez, \\ Joseph M. Hellerstein, Anthony D. Joseph, Aditya G. Parameswaran \\ UC Berkeley}

\maketitle

\section{Introduction}


During the course of Machine Learning (ML) model development,
a critical first step is {\em data validation},
ensuring that the data meets acceptable standards
necessary for input into ML training procedures.
Data validation involves various sub-tasks,
including 
{\em data preparation}:
transforming the data into a structured 
form suitable for the desired end-goal,
and  
{\em data cleaning:}
inspecting and fixing potential sources of errors.
These validation steps of data preparation and cleaning
are essential even if the eventual goal
is simply exploratory data analysis
as opposed to ML model development---in both
cases, the quality of the eventual
end-result, be it models or insights,
are highly tied to these steps.
This data validation process is highly exploratory and iterative, 
as the data scientist often 
starts off with a limited understanding 
of the data content and quality.
Data scientists therefore 
perform data validation through {\em incremental trial-and-error},
with the goals evolving over time: 
they make a change,
inspect the result (often just a sample) 
to see if it has improved or ``enriched'' 
the dataset in some way,
e.g., by removing outliers or filling in NULL values, 
expanding out a nested representation
to a flat relational one, or pivoting to organize
the dataset in a different manner 
more aligned with the analysis
goals.

To support this iterative process of trial-and-error,
data scientists often use powerful data analysis
libraries such as Pandas~\cite{pandas-api} within computational notebooks,
such as Jupyter or Google Colab~\cite{kluyver2016jupyter, colab}.
Pandas supports a rich set of incrementally specified 
operators atop a tolerant
dataframe-based data model, drawn from relational algebra,
linear algebra, and spreadsheets~\cite{petersohn13towards} 
embedded within a traditional imperative programming language, Python. 
While the use of dataframe libraries on computational notebooks
is a powerful solution for data validation on small datasets, 
this approach starts to break down on larger datasets~\cite{petersohn13towards}, with many operations requiring users to wait for
unacceptably long periods, breaking flow. 
Currently, this challenge may be overcome by either switching to a \emph{distributed dataframe system} (such as Dask~\cite{Dask} and Modin~\cite{Modin}), which introduces setup overhead and potential incompatibilities with the user's current workflow, or by users manually optimizing their queries, which is a daunting task as pandas has over 200 dataframe operations.
We identify two key opportunities for improving the interactive
user experience \textit{without requiring changes to user behavior}: 
\begin{itemize}
	\item Users often do not want to inspect the entire results of every single step.
	\item Users spend time thinking about what action to perform next. 
\end{itemize}
Unfortunately, at present, every cell (the unit of execution
in a notebook) issued by the
user is executed verbatim immediately, 
with the user waiting until execution is complete
to begin their next step. Moreover, the system is idle 
during {\em \thinktime}, i.e., when users are thinking
about their next step or writing code.
Fundamentally, {\em specification} 
(how the user writes the query)
and {\em execution} (what the system executes)
are tightly coupled. 

In this paper, we outline our initial insights and results
towards optimizing dataframe queries for interactive workloads
by {\em decoupling specification and execution}.
In particular, dataframe queries are not executed immediately,
unless the user intends to inspect the results,
but are deferred to be computed during \thinktime. 
We distinguish operators that produce results that users inspect, what we call \emph{interactions}, from those that do not.
We can then use program slicing to quickly determine what code is
{\em critical} in that it influences the interactions, i.e., what the user intends to see immediately,
and what is {\em non-critical}, in that it can be computed in
the background during think-time to speed up future interactions.
For the critical portions, we further identify
if it can be rewritten in ways that allows us to improve
interactivity further. For example, identifying
that users often only examine the first or 
last few rows/columns
of the result allows us to compute this
as part of the critical portion 
and defer the rest to the non-critical portion.
For the non-critical portions, by deferring the execution of the non-critical
portions, we can perform more holistic query planning
and optimization.
Moreover, we may 
also {\em speculatively} compute
other results that may prove useful in subsequent processing.
We call our framework {\em opportunistic evaluation},
preserving the benefits of eager evaluation (in that critical portions
are prioritized), and lazy or deferred evaluation (in that non-critical
portions are deferred for later computation). 
This paper builds on our prior vision~\cite{petersohn13towards},
wherein we outline our first steps towards establishing 
a formal framework for reasoning about dataframe optimization systematically.

\section{Background and Motivation}

\subsection{Key Concepts}
Users author dataframe queries in Jupyter notebooks,  comprising code cells and output from executing these code cells. Figure~\ref{fig:orig_files} shows an example notebook containing dataframe queries. Each code cell contains one or more queries and sometimes ends with a query that outputs results. In Figure~\ref{fig:orig_files}, every cell ends in a query (namely, \code{df1.describe()}, \code{df1.head()}, and \code{df2.describe()}) that outputs results.
Dataframe queries are comprised of operators such as \code{apply} (applying a user defined function on rows/columns), \code{describe} (compute and show summary statistics), and \code{head} (retrieve the top $K$ rows of the dataframe). Operators such as \code{head} and \code{describe}, or simply the dataframe variable itself, are used for inspecting intermediate results. We call these operators \textbf{\textit{interactions}}. 
Users construct queries incrementally by introducing interactions to verify intermediate results. An interaction usually depends on only a subset of the operators specified before it. 
For example, \code{df1.describe()} in Figure~\ref{fig:orig_files} depends only on 
\code{df1 = pd.read\_csv("small\_file")}
but not
\code{df2 = pd.read\_csv("LARGE\_FILE")}.
We call the set of dependencies of an interaction the \textbf{\textit{interaction critical path}}. To show the results of a particular interaction, the operators not on its interaction critical path do not need to be executed even if they were specified before the interaction.

After an interaction, users spend time inspecting the output and authoring new queries based on the output. We call the time between the display of the output and the submission of the next query \textbf{\textit{\thinktime}}, during which the CPU is idle (assuming there are no other processes running on the same server) while the user inspects intermediate results and authors new queries.
We propose \textbf{\textit{opportunistic evaluation}}, an optimization framework that leverages this \thinktime to reduce interactive latency. 
In this framework, the execution of operators that are not on interaction critical paths, which we call \textit{non-critical operators}, are deferred to being evaluated asynchronously during \thinktime to speed up future interactions.

\subsection{Motivating Scenarios and Example Optimizations}
\label{sec:tasks}
To better illustrate the optimization potential of opportunistic evaluation, we present two typical data analysis scenarios that could benefit from asynchronous execution of queries during \thinktime to minimize interactive latency. 
While the user's program remains the same, we illustrate the modifications to the execution plan that highlights the transformations made.

\subsubsection{Interaction-based Reordering}
\begin{figure}[h!]
    \centering
    \begin{subfigure}{0.4\textwidth}
    \vspace{-3pt}
         \centering
         \includegraphics[width=\textwidth]{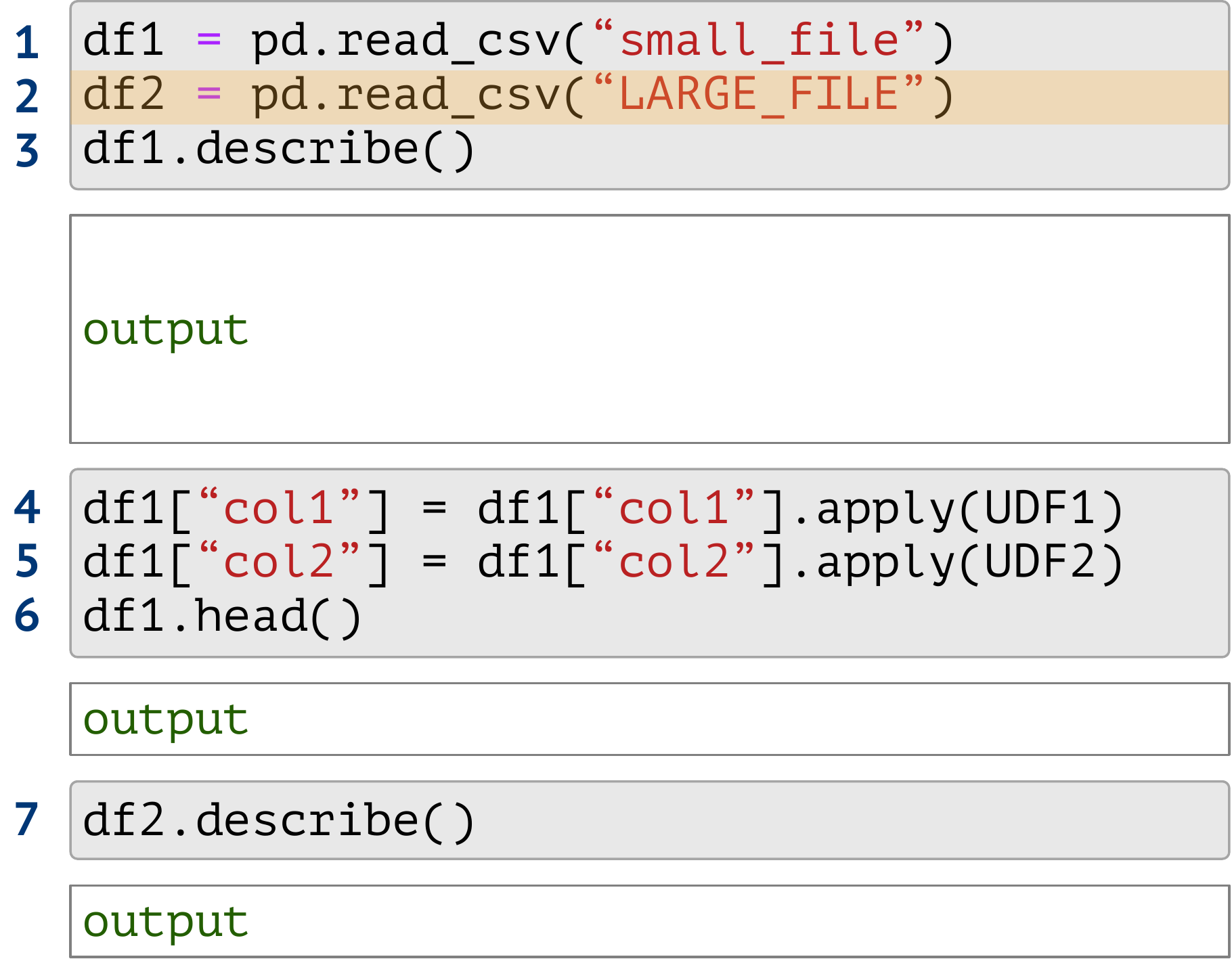}
         \caption{Original program where user has to wait for both files to load before viewing any.}
         \label{fig:orig_files}
    \end{subfigure}
    \hspace{10pt}
    \begin{subfigure}{0.44\textwidth}
         \centering
         \includegraphics[width=\textwidth]{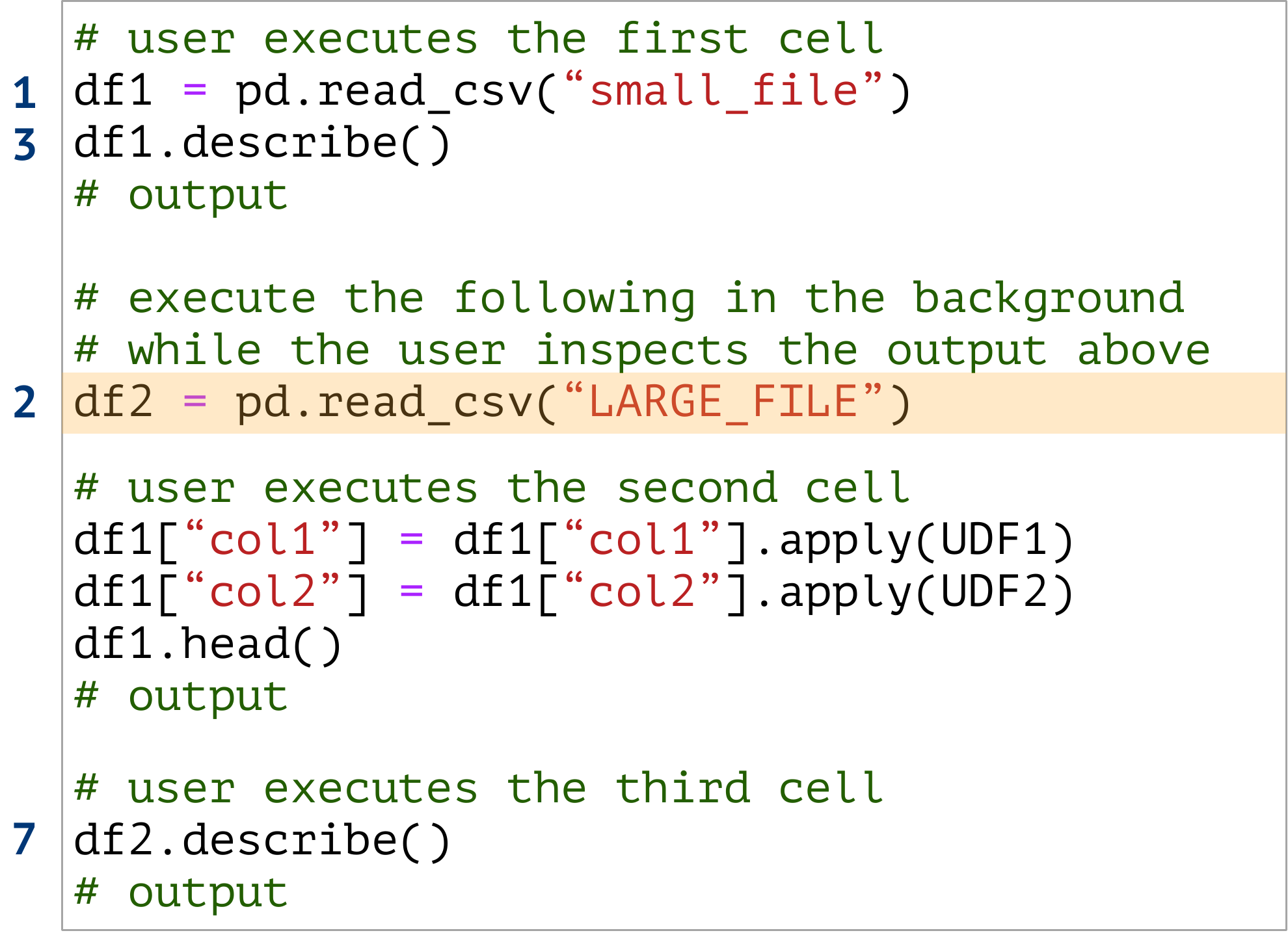}
         \caption{Optimized program where the user can view the smaller file first while the other loads.}
         \label{fig:opt_files}
    \end{subfigure}
    \vspace{-8pt}
    \caption{
    Example program transformation involving operator reordering.
    }
    \label{fig:files}
    \vspace{-8mm}
\end{figure}

Consider a common workflow of analyzing multiple data files, shown in Figure~\ref{fig:orig_files}.
The user, Sam, executes the first cell, which loads both of the files, and is forced to wait for \emph{both} to finish loading before she can interact with \emph{either} of the dataframes. 
To reduce the interactive latency (as perceived by the user), we could conceptually re-order the code to optimize for the immediate output.  As shown in Figure~\ref{fig:opt_files}, the re-ordered program defers loading the large file to after the interaction, \code{df1.describe()}, obviating the need to wait for the large file to load into \code{df2} before Sam can start inspecting the content of the small file.
To further reduce the interactive latency, the system could load \code{df2} while Sam is viewing the results of \code{df1.describe()}. This way, the time-consuming process of loading the large file is completed during Sam's \thinktime, thus reducing the latency for interacting with \code{df2}.

\subsubsection{Prioritizing Partial Results}
\begin{figure}[h!]
     \centering
     \begin{subfigure}[b]{0.4\textwidth}
         \centering
         \includegraphics[width=\textwidth]{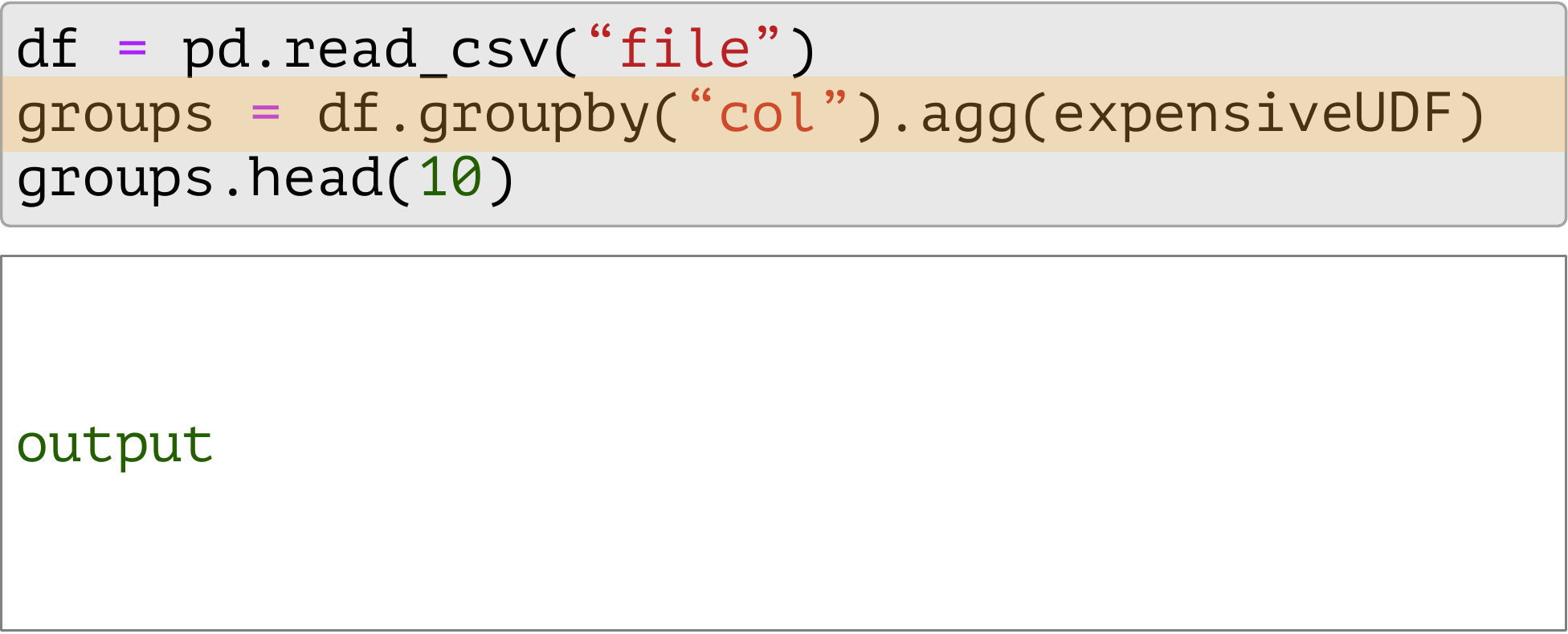}
         \caption{Original program where the user has to wait for an expensive UDFs to fully compute.}
         \label{fig:orig_partial}
     \end{subfigure}
     \hfill
     \begin{subfigure}[b]{0.55\textwidth}
         \centering
         \includegraphics[width=\textwidth]{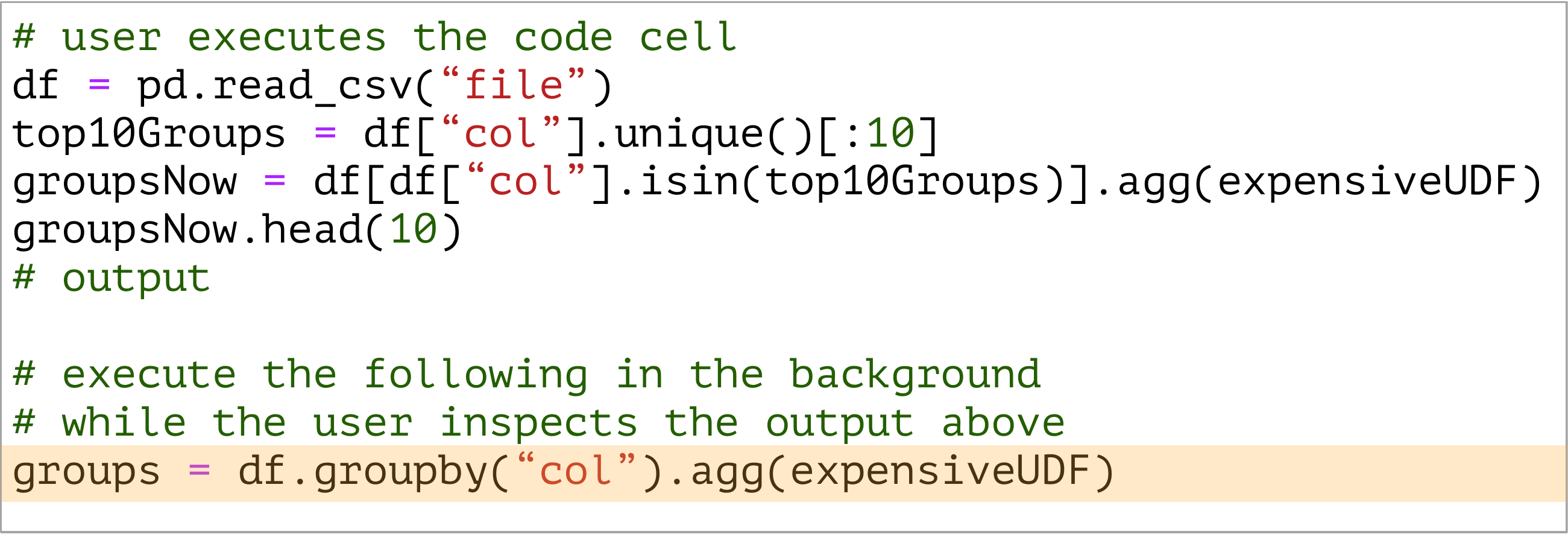}
         \caption{Optimized program where the user can view a partial result sooner.}
         \label{fig:opt_partial}
     \end{subfigure}
     \vspace{-8pt}
        \caption{Program transformation involving predicate pushdown.
        }
        \label{fig:partial}
    \vspace{-5mm}
\end{figure}

For any large dataframes, users can only inspect a handful of rows at a time.  However the current evaluation mechanism requires \emph{all} the rows to be evaluated.  Expensive queries such as those involving user-defined functions (UDFs) could take a long time to fully compute, as shown in Figure~\ref{fig:orig_partial}.

To reduce interactive latency, one can prioritize computation of only the portion of the dataframe inspected.  This method is essentially an application of \textit{predicate pushdown}, a standard technique from database query optimization.
Figure~\ref{fig:opt_partial} provides an example transformation for the particular operator, \code{groupby}.
While the first cell prioritizes the computation of the inspected rows, the user may still need the result of the entire computation, which is scheduled to be computed later while Sam is still reading the result of the previous cell, \code{groupNow.head(10)}, i.e. the \thinktime.
A noteworthy attribute of dataframes is row and column equivalence~\cite{petersohn13towards}, which means that
predicate pushdown can also happen when projecting columns as well.

\section{Assessment of Opportunities with Notebook Execution Traces}
\label{sec:formative}
To assess the size of opportunity for our aforementioned optimizations to reduce interactive latency in computational notebooks, we evaluate two real world notebook corpora.

One corpus is collected from students in the \textbf{Data 100} class offered at UC Berkeley. Data 100 is an intermediate data science course offered at the undergraduate level, covering topics on tools and methods for data analysis and machine learning.  This corpus contains 210 notebooks across four different assignments, complete with the \emph{history} of cell execution content and completion times captured by instrumenting a custom Jupyter extension.

We also collected Jupyter notebooks from \textbf{Github} comprising a more diverse group of users than Data 100.  Jupyter’s IPython kernel stores the code corresponding to each individual cell executions in a local \code{history.sqlite} file\footnote{\url{https://ipython.readthedocs.io/en/stable/api/generated/IPython.core.history.html}}.  We used 429 notebook execution histories that Macke et al.~\cite{mackefine} scraped from Github that also contained pandas operations. 

To assess optimization opportunities, we first quantify \thinktime between cell executions, and then evaluate the prevalence of the code patterns discussed in Section~\ref{sec:tasks}.

\subsection{Think-Time Opportunities}
\label{sec:thinktime}
\begin{figure}[h!]
     \centering
     \begin{subfigure}[b]{0.4\textwidth}
         \centering
         \includegraphics[width=\textwidth]{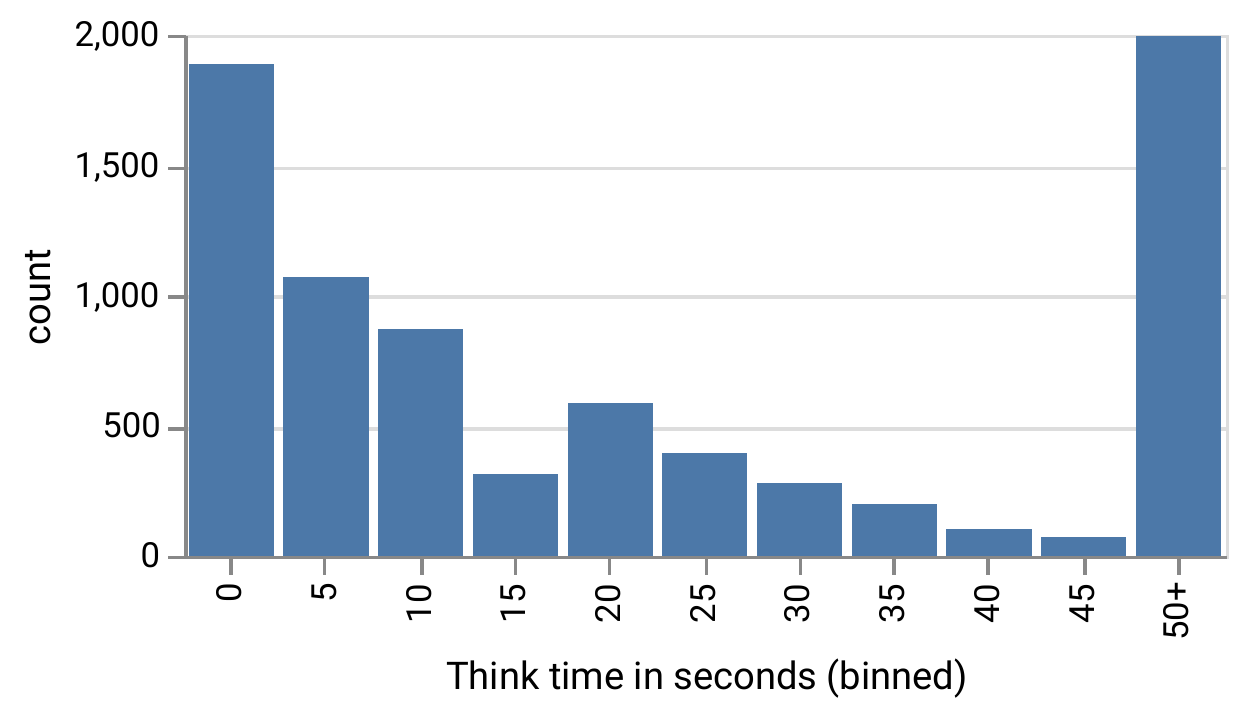}
         \caption{\Thinktime between cell executions.}
         \label{fig:think_cell}
     \end{subfigure}
    \hfill
     \begin{subfigure}[b]{0.4\textwidth}
         \centering
         \includegraphics[width=\textwidth]{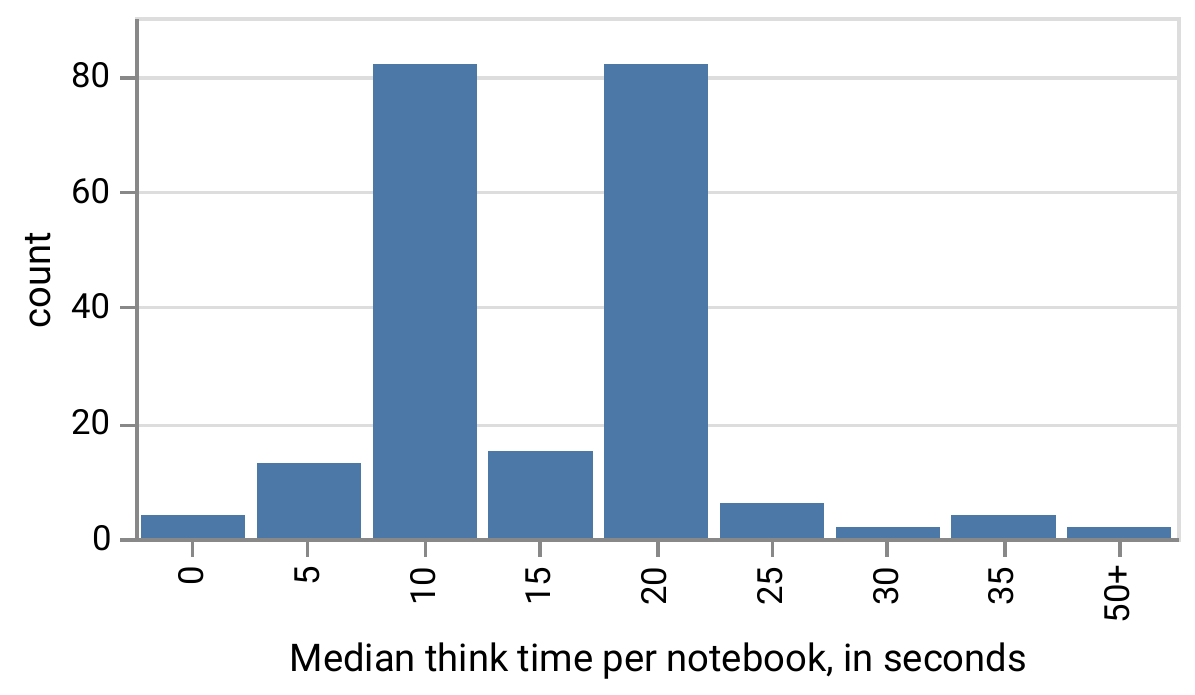}
         \caption{Median \thinktime per notebook across cells.}
         \label{fig:think_nb}
     \end{subfigure}
        \caption{\Thinktime the average ``\thinktime'' between cell executions and the average \thinktime per notebook.}
        \label{fig:think}
\end{figure}

Our proposed opportunistic evaluation framework takes advantage of user \thinktime to asynchronously process non-critical operators to reduce the latency of future interactions.
To quantify \thinktime, we measure the time lapsed between the completion of a cell execution and the start of the next cell execution using the timestamps in the cell execution and completion records, as collected by our Jupyter notebook extension.
Note that the \thinktime statistics are collected only on the Data 100 corpus, as the timestamp information is not available in the Github corpus.
Figure~\ref{fig:think_cell} shows the distribution of \thinktime intervals, measured in seconds, between consecutive cell executions across all notebooks, while Figure~\ref{fig:think_nb} shows the distribution of the median \thinktime intervals, measured in seconds, within each notebook.  
We removed automatic cell re-execution (``run all'') from the dataset.
We can see that while there are many cells that were executed quickly, there exist cells that had ample \thinktime---the 75th percentile \thinktime is 23 seconds. 

\subsection{Program Transformation Opportunities}

\begin{figure}[h!]
     \centering
    \begin{subfigure}[b]{0.4\textwidth}
         \centering
         \includegraphics[width=\textwidth]{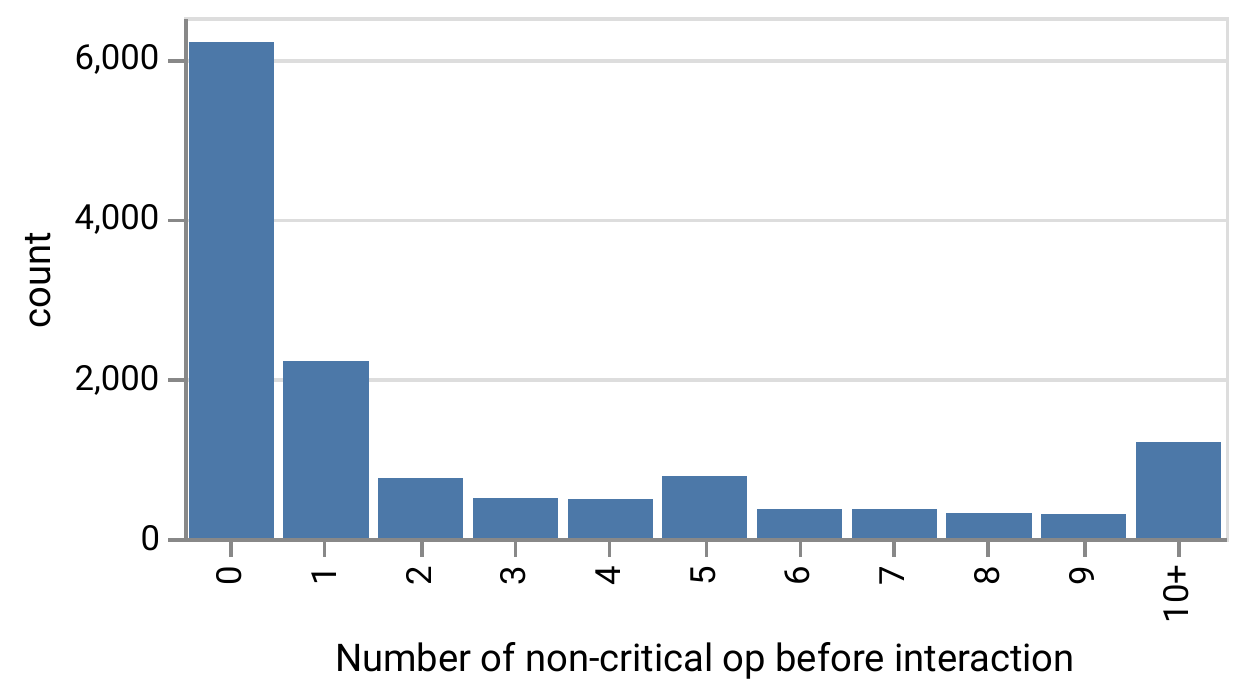}
         \caption{Data 100: $\mu$ = 4, $\sigma$ = 5}
         \label{fig:ds_nooo}
     \end{subfigure}
     \hfill
     \begin{subfigure}[b]{0.4\textwidth}
         \centering
         \includegraphics[width=\textwidth]{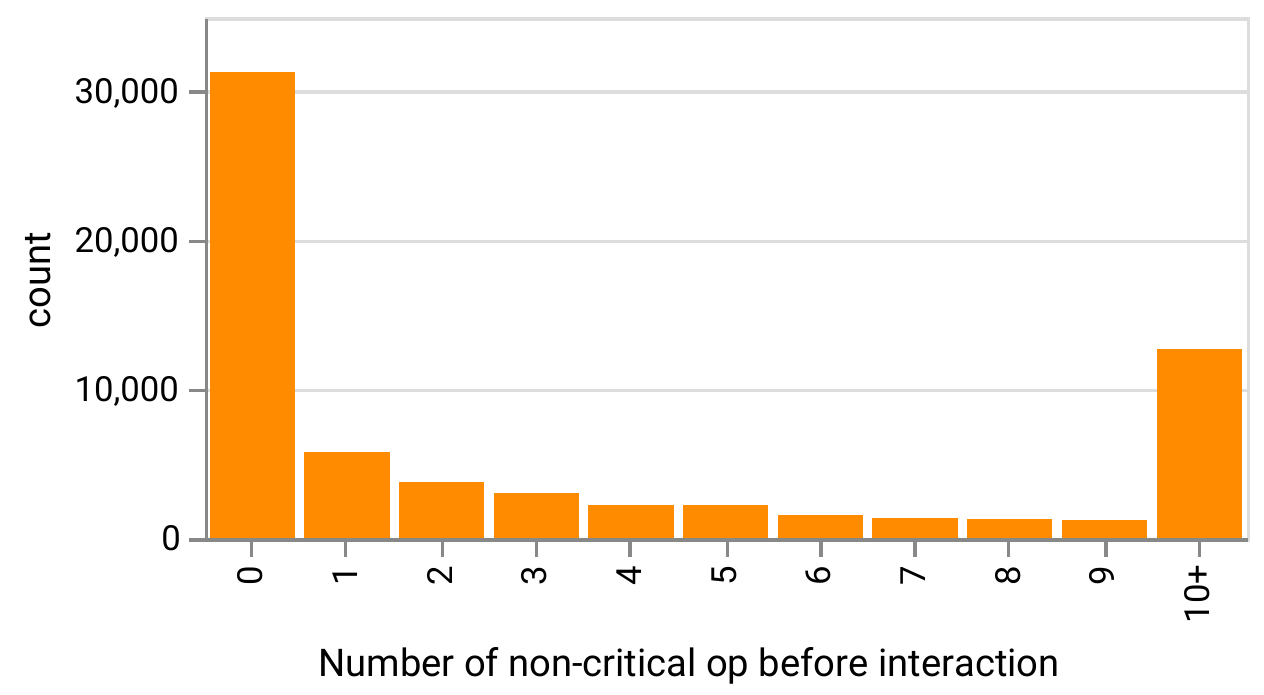}
         \caption{Github: $\mu$ = 7, $\sigma$ = 11}
         \label{fig:gh_nooo}
     \end{subfigure}
     \caption{Number of non-critical operators before interactions.}
    \label{fig:nooo}
    \vspace{-5mm}
\end{figure}

\topic{Interaction-Based Reordering} To assess the opportunities to apply operator reordering to prioritize interactions, we evaluate the number of non-critical operators specified before each interaction.  We use the operator DAG, to be described in Section~\ref{sec:dag}, to determine the dependencies of an interaction and count the number of operators that are not dependencies, i.e., non-critical operators, specified above the interaction.
Figure~\ref{fig:nooo} shows the distributions for the two datasets.
In both cases, non-critical operators present a major opportunity: the Data 100 and Github corpus have, respectively, 54\% and 42\% interactions with non-critical operators.

\begin{figure}[h!]
     \centering
     \begin{subfigure}[b]{0.4\textwidth}
         \centering
         \includegraphics[width=\textwidth]{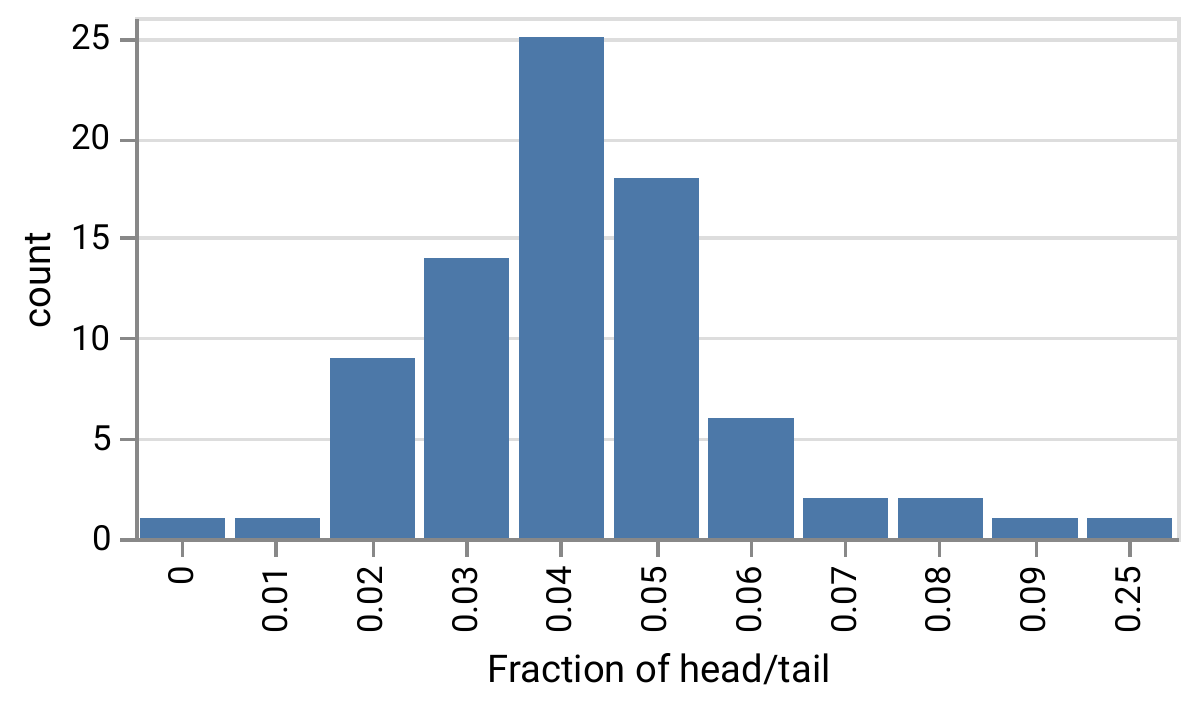}
         \caption{Data 100: $\mu$ = 0.04, $\sigma$ = 0.028}
         \label{fig:ds_nht}
     \end{subfigure}
     \hfill
     \begin{subfigure}[b]{0.4\textwidth}
         \centering
         \includegraphics[width=\textwidth]{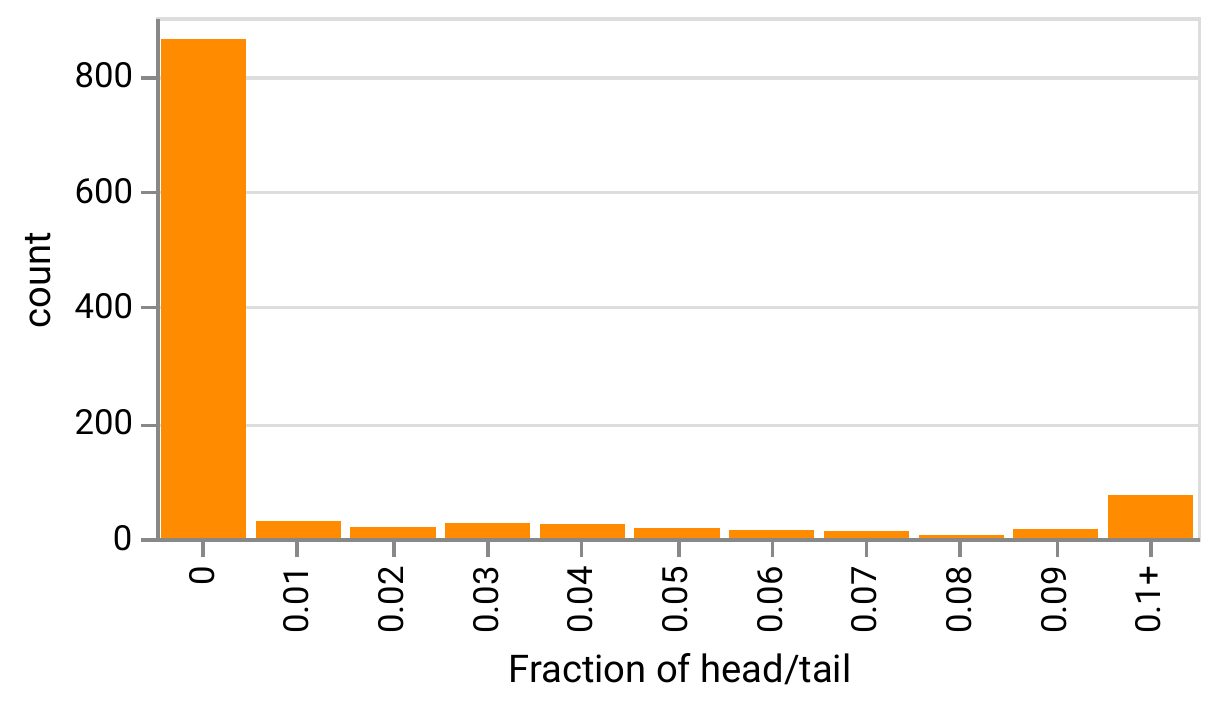}
         \caption{Github: $\mu$ = 0.11, $\sigma$ = 0.21}
         \label{fig:gh_nht}
     \end{subfigure}
     \caption{Stats for head/tail interactions used in each notebook.}
     \vspace{-3mm}
\end{figure}

\begin{figure}[h!]
     \centering
     \begin{subfigure}[b]{0.4\textwidth}
         \centering
         \includegraphics[width=\textwidth]{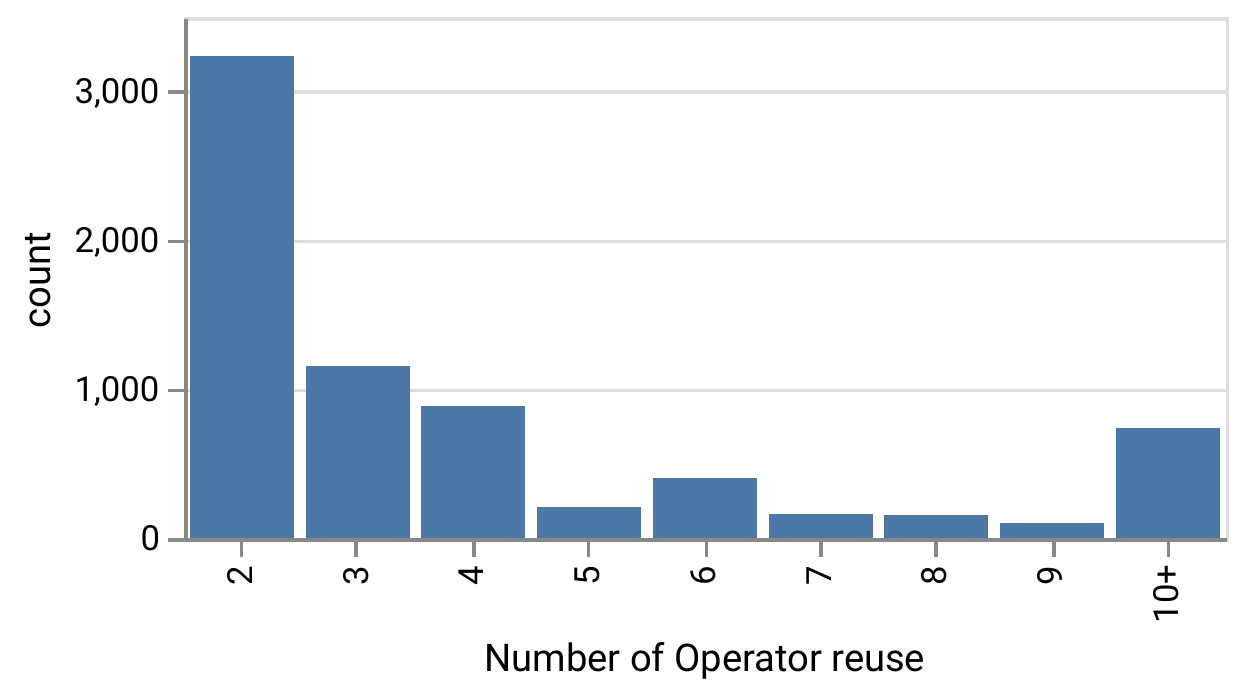}
         \caption{Data 100: $\mu$ = 5, $\eta$ = 3, $\sigma$ = 8}
         \label{fig:ds_nre}
     \end{subfigure}
     \hfill
     \begin{subfigure}[b]{0.4\textwidth}
         \centering
         \includegraphics[width=\textwidth]{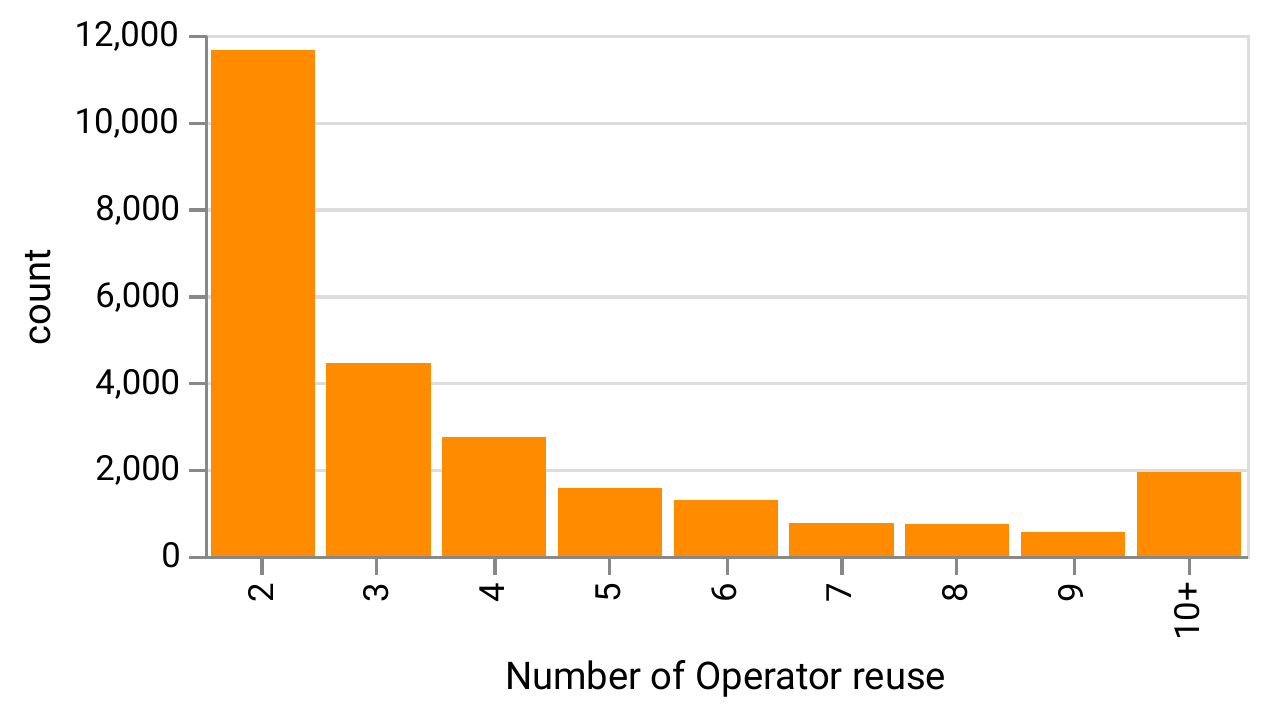}
         \caption{Github: $\mu$ = 7, $\eta$ = 3, $\sigma$ = 14}
         \label{fig:gh_nre}
     \end{subfigure}
    \caption{Distribution of number of operators that can benefit from reuse.}
    \label{fig:reuse}
    \vspace{-5mm}
\end{figure}

\topic{Prioritizing Partial Results}  The optimization for prioritizing partial results via predicate pushdown can be applied effectively to many cases when predicates are involved in queries with multiple operators.  The most common predicates in the dataframe setting are \code{head()} and \code{tail()}, which show the top and bottom $K$ rows of the dataframe, respectively. 
Figure~\ref{fig:gh_nht} and Figure~\ref{fig:ds_nht} show the distribution of the fraction of interactions that are either \code{head} or \code{tail} in each notebook. 
We see that partial results views are much more common in the GitHub dataset than Data 100. This could be due to the fact that users on GitHub tend to keep the cell output area short for better rendering of the notebook by Github, but further studies are needed to corroborate this hypothesis. Lastly, partial views are not nearly as prevalent as non-critical operators before an interaction, accounting only for $<20\%$ of the interactions. 

\topic{Reuse of Intermediate Results} 
Since dataframe queries are incrementally constructed, with subsequent queries building on top of previous ones, another common query optimization technique that is applicable is caching these intermediate results.
To assess the opportunities to speed up queries by caching, we evaluate the number of times an operator is shared by different interactions but not stored as a variable by the user.
Ideally, we would also have the execution times of the individual operators, which is not possible without a full replay.  We present an initial analysis that only assesses the \emph{existence} of reuse opportunities, as shown
in Figure~\ref{fig:gh_nre} and Figure~\ref{fig:ds_nre}. Both the Data 100 and Github datasets have a median of 3 operators that can benefit from reuse.

\vspace{2pt}
\noindent Of the types of optimizations explored, operator reordering appears to be the most common.
Thus, we focus our initial explorations of opportunistic evaluation on operator reordering for asynchronous execution during \thinktime, while supporting preemption to interrupt asynchronous execution and prioritize interaction.

\section{System Architecture}
\label{sec:arch}
In this section, we introduce the system architecture for implementing our
opportunistic evaluation framework for dataframe query optimization within Jupyter notebooks. 
At a high level, we create a custom Jupyter Kernel to intercept dataframe queries in order to defer, schedule, and optimize them transparently. The query execution engine uses an operator DAG representation for scheduling and optimizing queries and caching results, and is responsible for scheduling asynchronous query executions during \thinktime. When new interactions arrive, the execution of non-critical operators is preempted and partial results are cached to resume execution during the next \thinktime. A garbage collector periodically uncaches results corresponding to the DAG nodes to avoid memory bloat based on the likelihood of reuse.

\subsection{Kernel Instrumentation}
\begin{figure}
    \centering
    \includegraphics[width=0.7\textwidth]{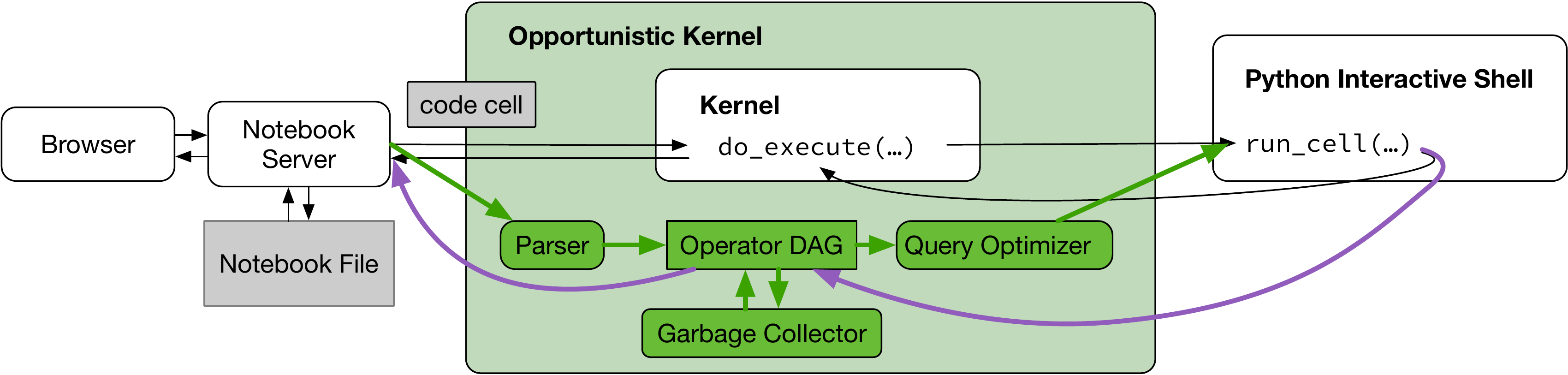}
    \caption{Opportunistic Evaluation Kernel Architecture.}
    \label{fig:workflow}
    \vspace{-5mm}
\end{figure}

Figure~\ref{fig:workflow} illustrates the round-trip communication between the Jupyter front-end and the Python interactive shell. The black arrows indicate how communication is routed normally in Jupyter, whereas the green and purple arrows indicate how we augment the Jupyter Kernel to enable opportunistic evaluation. First, when the code is passed from the front-end to the kernel, it is intercepted by the custom kernel that we created by wrapping the standard Jupyter kernel. As shown in the green box, the code is passed to a parser that generates a custom intermediate representation, the operator DAG. The operator DAG is then passed to the query optimizer to create a physical plan for the query to be executed. This plan in then passed to the Python interactive shell for execution. When the shell returns the result after execution, the result is intercepted by the custom kernel to augment the operator DAG with runtime statistics as well as partial results to be used by future queries, and the query results are passed back to the notebook server, as indicated by the purple arrows.

\subsection{Intermediate Representation: Operator DAG}
\label{sec:dag}
\begin{figure}
    \centering
    \includegraphics[width=0.8\textwidth]{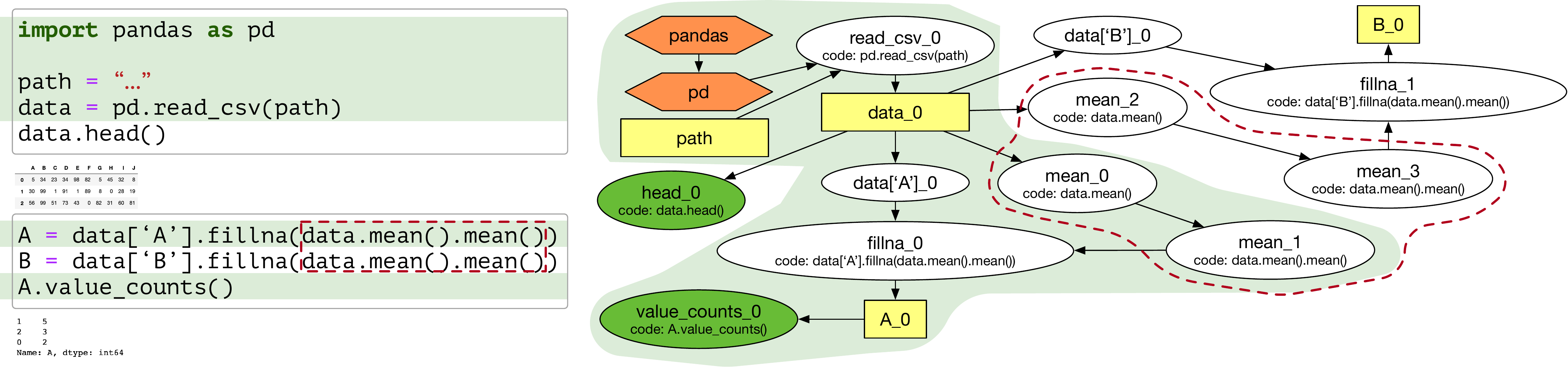}
    \caption{Example Code Snippet and Operator DAG.}
    \label{fig:op_dag}
    \vspace{-5mm}
\end{figure}

Figure~\ref{fig:op_dag} shows an example operator DAG constructed from the code snippet on the left. The orange hexagons are imports, yellow boxes are variables, ovals are operators, where green ovals are interactions. The operator DAG is automatically constructed by analyzing the abstract syntax tree of the code, in the parser component in Figure~\ref{fig:workflow}. We adopt the static single assignment form in our operator DAG node naming convention to avoid ambiguity of operator references, as the same operator can be invoked many times, either on the same or different dataframes. 
In the case that the operator DAG contains non-dataframe operators, we can simply project out the irrelevant operators by keeping only the nodes that are weakly connected to the \code{pandas} import node. 

To see how the operator DAG can be used for optimization, consider two simple use cases:

\topic{Critical path identification} To identify the critical path to the interaction \code{A.value\_counts()}, we can simply start at the corresponding node and traverse the DAG backwards to find all dependencies. Following this procedure, we would collect all nodes in the green region as the critical path to \code{A.value\_counts()} (corresponding statements are highlighted in green on the left), slicing out the operators associated with the statement \code{B = data[‘B’].fillna(data.mean().mean())}, which does not need to be computed for the interaction. 

\topic{Identifying repeated computation} Note that \code{data.mean().mean()} is a common subexpression in both \code{A} and \code{B}; recognizing this allows us to cache and reuse the result for \code{data.mean().mean()}, which is expensive since it requires visiting every element in the dataframe. We assume that operators are \textit{idempotent}, i.e., calling the same operators on the same inputs would always produce the same results. Thus, descendants with identical code would contain the same results. Based on this assumption, we eliminate common subexpressions by starting at the root nodes and traversing the graph breadth first, merging any descendants with identical code. We then proceed to the descendants of the descendants and carry out the same procedure until the leaf nodes are reached. Following this procedure, we would merge \code{mean\_0} with \code{mean\_2} and \code{mean\_1} with \code{mean\_3} in the red dotted region in Figure~\ref{fig:op_dag}.

We will discuss more optimizations in Section~\ref{sec:opt}.

\subsection{Operator Execution \& Garbage Collector}
When a notebook cell is executed, the opportunistic kernel first parses the code in the cell to add operators to the operator DAG described above. The DAG is then passed to the query optimizer, which will either immediately kick off the execution of interaction critical paths, if they are present in the DAG, or consider all the non-critical operators to determine what to execute next. We discuss optimizations for non-critical operators in Section~\ref{sec:non-critical}.

After the last interaction is executed and the results are returned, the query optimizer will continue executing operators asynchronously until the entire DAG is executed.
In the event that an interaction arrives while a non-critical operator is executing, we preempt the execution of the non-critical operator to avoid delaying the execution of the interaction critical path. We discuss optimizations for supporting effective preemption in Section~\ref{sec:critical}.

While the kernel executes operators, a garbage collector (GC) is working in the background to uncache results in the operator DAG to control memory consumption. A GC event is triggered when memory consumption is above 80\% of the maximum memory allocated to the kernel, at which point the GC inspects the operator DAG to uncache the set of operator results that are the least likely to speed up future queries. We discuss cache management in Section~\ref{sec:non-critical}.

\section{Optimization Framework}
\label{sec:opt}
The opportunistic evaluation framework optimizes for interactive latency by deferring to \thinktime the execution of operators that do not support interactions. The previous section describes how we use simple program analysis to identify the interaction critical path that must be executed to produce the results for an interaction. In this section, we discuss optimizations for minimizing the latency of a given interaction in Section~\ref{sec:critical} and optimizations for minimizing the latency of future interactions by leveraging \thinktime in Section~\ref{sec:non-critical}. We discuss how to model user behavior to anticipate future interactions in Section~\ref{sec:user}.

\subsection{Optimizing Current Interactions}
\label{sec:critical}

Given an interaction critical path, we can apply standard database optimizations for single queries to optimize interactive latency. For example, if the interaction operator is \code{head} (i.e., examining the first $K$ rows), we can perform predicate pushdown to compute only part of the interaction critical path that leads to the top $K$ rows in the final dataframe. The rest can be computed during \thinktime in anticipation of future interactions.

The main challenge for optimizing interactive latency in opportunistic evaluation is the ability to effectively preempt the execution of non-critical operators. This preemption ensures that we avoid increasing the interactive latency due to irrelevant computation. 
The current implementation of various operators within pandas and other dataframe libraries often involves calling lower-level libraries that cannot be interrupted during their execution. In such cases, the only way to preempt non-critical operators is to abort their execution completely, potentially wasting a great deal of progress. We propose to overcome this challenge by partitioning the dataframe so that preemptions lead to, in the worst case scenario, only loss of the progress on the current partition.

\topic{Dataframe partitioning}
Partitioning the dataframe in the opportunistic evaluation setting involves navigating the trade-off between the increase in future interactive latencies due to loss of progress during preemption and the reduction in operator latency due to missed holistic optimizations on the entire dataframe. In the setting where interactions are sparse, a single partition maximizes the benefit of holistic optimization while losing progress on the entire operator only occasionally due to preemption. On the other hand, if interactions are frequent and erratic, a large number of small partitions ensures progress checkpointing, at the expense of longer total execution time across all partitions. Thus, the optimal partitioning strategy is highly dependent on user behavior. We discuss how to model user behavior in Section~\ref{sec:user}.

Without a high-fidelity user interaction model, we can create unevenly sized partitions to handle the variability in the arrival rate of interactions. First, we create small partitions for the top and bottom $K$ rows in the dataframe not only to handle the rapid succession of interactions but also to support partial-result queries involving \code{head} and \code{tail} that are prevalent in interactive dataframe workloads. Then, for the middle section of the dataframe, the partitions can reflect the distribution of \thinktime such that the partition sizes are smaller at intervals where interactions are likely to be issued. For example, if the median \thinktime is 20s and the operator's estimated execution time is 40s, it might be desirable to have smaller partitions after ~50\% of the rows have been processed. 

The above strategy assumes sequential processing of every row in the dataframe. If, instead, the prevalent workload is working with a select subset of rows, then it is more effective to partition based on the value of the attributes that are commonly used for selection. Of course, partitioning is not necessary if computation started during \thinktime does not block computation for supporting interactions. 

Note that another important consideration in generating partial results is the selectivity
of the underlying operators and whether they are blocking operators. For the former, we may need
to employ a much larger partition simply
to generate $K$ results.
For the latter, we may need to prioritize the
generation of the aggregates corresponding to
the groups in the top or bottom $K$ (in the case
of group-by), or to employ algorithms 
that prioritize the generation of the $K$ first
sorted results (in the case of sorting).
In either case, the problem becomes a lot more 
challenging. 

\subsection{Optimizing Future Interactions Leveraging Think Time}
\label{sec:non-critical}

\stitle{Non-critical Operator scheduling.} We now discuss scheduling non-critical operators. 
Recall that these operators are organized in a DAG built from queries. 
The job of our scheduler is to decide which \textit{source} operators to execute. 
Source operators in the DAG are those whose precedent operators 
do not exist or are already executed. 
We assume an equal probability of users selecting any operator in the DAG to extend with an interaction. 

The scheduler is optimized to reduce the interaction latency; we introduce
the notion of an operator's \textit{delivery cost} as the proxy for it. 
If an operator has not been executed yet, its delivery cost is the cost of executing the operator along with all of its unexecuted predecessors. 
Otherwise, the delivery cost is zero.
Our scheduler prioritizes scheduling the source operator that can reduce the delivery cost across all operators the most. 
We define a utility function $U(s_i)$ to estimate the benefit of executing a source operator $s_i$.
This function, for a node $s_i$ 
is set to be the sum of the delivery cost 
for the source operator and all of its successors $D_i$:
\begin{equation}
\label{eq:utility}
U(s_i) = \sum_{j \in D_i}{c_j}
\end{equation}
where $c_j$ is the delivery cost for an operator $j$.
Our scheduler chooses to execute the one with the highest $U(s_i)$.
This metric prioritizes those operators that ``influence'' as many 
expensive downstream
operators as possible.

\stitle{Caching for reuse.}
When we are executing operators in the background, we store the result of each newly computed operator in memory. 
However, if the available memory (i.e., the memory budget) is not sufficient to store the new result, 
we need to recover enough memory by discarding materialized results of previously computed operators. 
If the discarded materialized results are needed by future operators, we will execute the corresponding operators to recompute them. 
Here, the optimization problem is to determine which materialized results should be discarded given the memory budget. 
Our system addresses this problem by systematically considering three aspects of a materialized result, denoted $r_i$: 
1) the chance of $r_i$ being reused, $p_i$, 
2) the cost of recomputing the materialized result, $k_i$, 
and 3) the amount of memory it consumes, $m_i$. 
We estimate $p_i$ by borrowing ideas from the LRU replacement algorithm. 
We maintain a counter $T$ to indicate the last time any materialized result is reused 
and each materialized result is associated with a variable $t_i$ that tracks the last time it is reused.
If one materialized result $r_i$ is reused, we increment the counter $T$ by one and set $t_i$ to $T$. 
We use the following formula to estimate $p_i$: 
\begin{equation}
p_i = \frac{1}{T + 1 - t_i}
\end{equation}
We see that the more recently a materialized result $r_i$ is reused, the higher $p_i$ is. 
We can use a cost model as in relational databases to estimate the recomputation cost $k_i$. 
We note that we do not always recompute an operator from scratch. 
Given that the other materialized results are in memory, 
our cost model estimates the recomputation cost by considering reusing existing materialized results. 
Therefore, we use the following utility function to decide which materialized result should be discarded.
\begin{equation}
O(r_i) = p_i \times \frac{m_i}{k_i}
\end{equation}
Here, $\frac{m_i}{k_i}$ represents the amount of memory we can spare per unit of recomputation cost to pay. 
The lower $\frac{m_i}{k_i}$ is, the more likely we discard $r_i$. 
Finally, our algorithm will discard the $r_i$ with the lowest $O(r_i)$ value.

\stitle{Speculative materialization.}
Our system not only considers caching results generated by users' programs, 
but also speculatively materializes and caches results that go beyond what users specify, to be used by future operators. 
One scenario we observed is that users intend to explore the data by changing the value of a filter 
repeatedly. In this case, we can materialize the intermediate output results before we apply the filter 
and when users modify the filter, we can reuse the saved results without computing them from scratch. 
The downside of this approach is that it can increase the latency of computing an interaction when the \thinktime is limited. 
Therefore, we enable this optimization only when users' predicted \thinktime of writing a new operator 
is larger than the time of materializing the intermediate states.

\subsection{Prediction of User Behavior}
\label{sec:user}
The accurate prediction of user behavior can greatly improve the efficacy of opportunistic evaluation. 
Specifically, we need to predict two types of user behavior: \thinktime 
and future interactions. 
Section~\ref{sec:thinktime} described some preliminary statistics that can be used to construct a prior distribution for \thinktime. As the system observes the user work, this distribution can be updated to better capture the behavior of the specific user, as we expect the distribution of \thinktime to vary greatly based on the dataset, task, user expertise, and other idiosyncrasies. These workload characteristics can be factored into the \thinktime model for more accurate prediction. This \thinktime model can be used by the optimizer to decide the size of dataframe partitions to minimize progress loss due to preemption or to schedule non-critical operators whose expected execution times are compatible with the \thinktime duration.

To predict future interactions, we can use the models from Yan et al.~\cite{yan2020auto}.
These models are trained on a large corpus of data science notebooks from Github. 
Since future interactions often build on existing operators, we can use the future interaction prediction model to estimate the probabilities of non-critical operators in the DAG leading to future interactions, which can be used by the scheduler to pick non-critical operators to execute next.
Let $p_j$ be the probability of the children of an operator $j$ being an interaction. We can incorporate $p_j$ into the utility function in Equation~\ref{eq:utility} to obtain the updated utility function:
\begin{equation}
    U_p(s_i) = \sum_{j \in D_i} {c_j \times p_j}
\end{equation}

Of course, the benefits of opportunistic evaluation can lead to modifications in user behavior. For example, without opportunistic evaluation, a conscientious user might self-optimize by avoiding specifying expensive non-critical operators before interactions, potentially at the cost of code readability. When self-optimization is no longer necessary when authoring queries, the user may choose to group similar operators for better code readability and maintenance, thus creating more opportunities for opportunistic evaluation optimizations. 

\section{Case Study}

\begin{figure}
    \centering
         \centering
         \includegraphics[width=\textwidth]{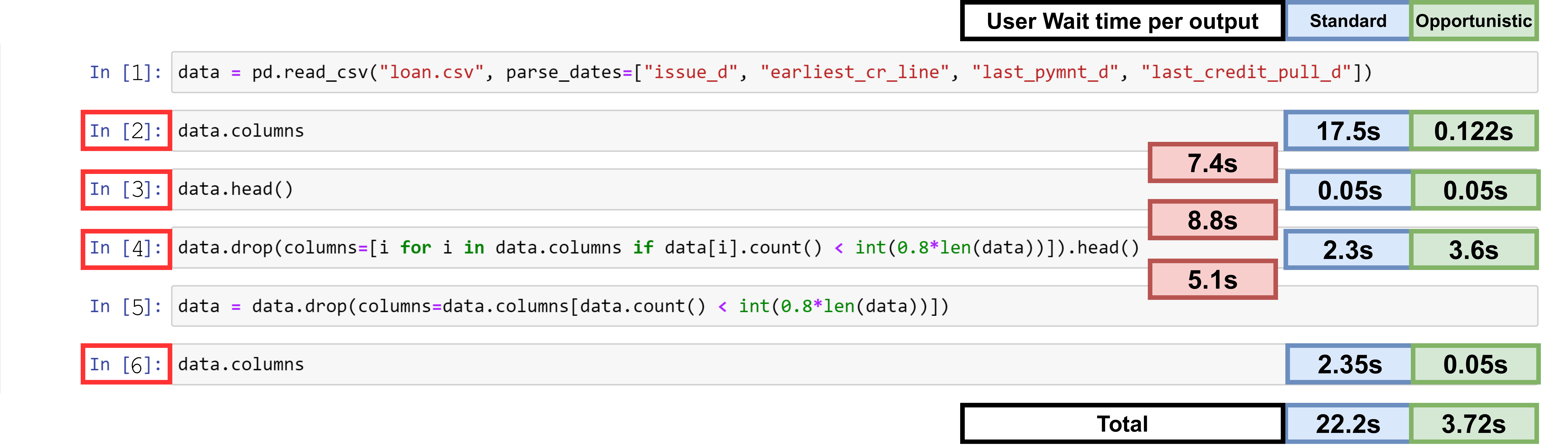}
         \caption{
         An example notebook. Cells that show an output are indicated with a red box.
         }
         \label{fig:case_study_before}
         \vspace{-5mm}
\end{figure}

In this section, we evaluate how opportunistic evaluation will impact the end user
through a case study. Figure~\ref{fig:case_study_before} shows an excerpt from
the original notebook, taken from a Kaggle competition (https://www.kaggle.com/c/home-credit-default-risk). 

In this case study, the data scientist first read in the file, and was forced to immediately
wait. Then, the user wanted to see the columns that exist in the dataset. This is
often done when the data scientist first encounters a dataset. They therefore
printed the first 5 lines with \code{data.head()}. This inspection is important for
data validation: the data scientist wanted to ensure that the data was parsed
correctly during data ingestion. After these two data validation steps, the data scientist
noticed that there were a significant number of \code{null} values in multiple
columns. 

The cell labeled \code{In[4]}
shows
how the data scientist solved the \code{null} values problem: they decided to drop any
column that does not have at least 80\% of its values present. Notice
that the data scientist first wanted to see what the results of the query
would look like before they executed it, so
they added a \code{.head()} to the end of the query that drops the columns.
Likely this was done during debugging, where many different, but similar queries
were attempted until the desired output was achieved. The query was then
repeated to overwrite the \code{data} variable. An important note here is that
the full dataset is lost at this point due to the overwriting of the \code{data}
variable. The data scientist will need to reread the file if they want access to 
the full dataset again. After dropping columns with less than 80\% of their values
present, the data scientist
double-checked their work by inspecting the \code{columns} of the overwritten
\code{data} dataframe.
Next, we evaluate the benefits of the opportunistic evaluation approach by determining the amount of
synchronous wait time saved by leveraging \thinktime.

To evaluate opportunistic evaluation in our case study, \thinktime was injected into
the notebook from the distribution presented in Figure~\ref{fig:think}.
We found that the time that the hypothetical data scientist spent waiting on computation was
almost none: the \code{read\_csv} phase took 18.5 seconds originally, but since
the output of the \code{columns} and \code{head} were prioritized, they were
displayed almost immediately (122ms). The data scientist then looked at the
two outputs from \code{columns} and \code{head} for a combined 16.2 seconds.
This means the data scientist synchronously waited on the \code{read\_csv} for approximately
1.3 seconds. Next, the user had to wait another 2.3 seconds for the columns
with less than 80\% of their values present to be dropped. Without opportunistic
evaluation, the user would have to pay this time twice, once to see the first
5 lines with \code{head} and again to see the \code{data.columns} output in 
cell \code{In[6]}.

\section{Related Work}
Recently, researchers and developers have begun to turn their attention to the optimization, 
usability, and scalability of dataframes as the community begins to recognize its important role in data exploration and analysis. 
Some of these issues are brought on by the increasingly complex API and ever-growing data sizes. 
Pandas itself has best practices for performance optimization embedded within
its user guide~\cite{pandas-opt}. However, enabling these optimizations often requires a change to the user's behavior.

Many open source systems attempt to provide improved performance or scalability for dataframes. Often, this means only supporting
dataframe functionalities that are simple to parallelize (e.g., Dask~\cite{Dask}),
or supporting only those operations which can be represented in SQL 
(e.g. Koalas~\cite{koalas} and SparkSQL~\cite{sparksql-func}). Our project, Modin~\cite{Modin}, is the only open source
system with an architecture that can support all dataframe operators.

In the research community, there are multiple notable papers that have tackled dataframe optimization through vastly different approaches. Sinthong et al. propose AFrame, a dataframe system implemented on top of AsterixDB by translating dataframe APIs into SQL++ queries that are supported by AsterixDB~\cite{sinthong2019aframe}. 
Another work by Yan et al. aims to accelerate EDA with dataframes by ``auto-suggesting'' data exploration operations~\cite{yan2020auto}. 
Their approach has achieved considerable success in predicting the operations that were actually carried out by users given an observed sequence of operations. More recently, Hagedorn et. al. designed a system for translating pandas operations to SQL and executing on existing RDBMSs~\cite{hagedornputting}. In a similar vein, Jindal et. al. built a system called Magpie for determining the optimal RDBMS to execute a given query~\cite{jindalmagpie}. Finally, Sioulas et. al. describe techniques for combining the techniques from recommendation systems to speculatively execute dataframe queries~\cite{sioulasaccelerating}. 

Our proposed approach draws upon a number of well established techniques from the systems, PL, and DB communities. Specifically, determining and manipulating the DAG of operators blends control flow and data flow analysis techniques from the PL community~\cite{cooper2011engineering}. 
The optimization of dataframe operators draws inspiration from battle-tested database approaches such as predicate pushdown, operator reordering, multi-query optimization, and materialized views~\cite{hellerstein2007architecture}, as well as compiler optimizations such as program slicing and common subexpression elimination. Furthermore, we borrow from the systems literature on task scheduling to take enable asynchronous execution of dataframe operators during think time.

\section{Conclusion \& Future Work}
We proposed opportunistic evaluation, a framework for accelerating interactions with dataframes. Interactive latency is critical for iterative, human-in-the-loop dataframe workloads for supporting data validation, both for ML and for EDA. Opportunistic evaluation significantly reduces interactive latency by 1) prioritizing computation directly relevant to the interactions and 2) leveraging \thinktime for asynchronous background computation for non-critical operators that might be relevant to future interactions. We have shown, through empirical analysis, that current user behavior presents ample opportunities for optimization, and the solutions we propose effectively harness such opportunities. 

While opportunistic evaluation addresses data validation prior to model training, data validation challenges are present in other parts of the end-to-end ML workflow. For example, after a trained model has been deployed, it is crucial to monitor and validate online data against the training data in order to detect data drift, both in terms of distribution shift and schema changes. A common practice to address data drift is to retrain the model on newly observed data, thus introducing data drift into the data pre-processing stage of the end-to-end ML workflow. Being able to adapt the data validation steps in a continuous deployment setting to unexpected data changes is an open challenge.

\bibliographystyle{abbrv}
{\small
\bibliography{main.bib}
}

\end{document}